\documentclass[aps,prc,twocolumn,superscriptaddress,showpacs]{revtex4-1}

\usepackage{graphicx}
\usepackage{multirow}
\usepackage{longtable}
\usepackage{tabularx}
\usepackage{lipsum}

\begin{document}


\title{Mirror nucleon-transfer reactions from $^{18}$Ne and $^{18}$O}


\author{F.~Flavigny}
\affiliation{Universit\'{e} de Caen Normandie, ENSICAEN, CNRS/IN2P3, LPC Caen UMR6534, F-14000 Caen, France}
\author{N.~Keeley}
\affiliation{National Centre for Nuclear Research, ul.\ Andrzeja So\l tana 7, 05-400 Otwock, Poland}
\author{A.~Gillibert}
\author{V.~Lapoux}
\affiliation{Universit\'{e}  Paris-Saclay, IRFU, CEA, F-91191 Gif-sur-Yvette, France}
\author{A.~Lemasson}
\affiliation{Grand Acc\'el\'erateur National d’Ions Lourds (GANIL), CEA/DRF-CNRS/IN2P3, Bvd Henri Becquerel, 14076 Caen, France}
\author{L.~Audirac}
\affiliation{Universit\'{e}  Paris-Saclay, IRFU, CEA, F-91191 Gif-sur-Yvette, France}
\author{B.~Bastin}
\affiliation{Grand Acc\'el\'erateur National d’Ions Lourds (GANIL), CEA/DRF-CNRS/IN2P3, Bvd Henri Becquerel, 14076 Caen, France}
\author{S.~Boissinot}
\affiliation{Universit\'{e}  Paris-Saclay, IRFU, CEA, F-91191 Gif-sur-Yvette, France}
\author{J.~Caccitti}
\affiliation{Grand Acc\'el\'erateur National d’Ions Lourds (GANIL), CEA/DRF-CNRS/IN2P3, Bvd Henri Becquerel, 14076 Caen, France}
\author{A.~Corsi}
\affiliation{Universit\'{e}  Paris-Saclay, IRFU, CEA, F-91191 Gif-sur-Yvette, France}
\author{S.~Damoy}
\affiliation{Grand Acc\'el\'erateur National d’Ions Lourds (GANIL), CEA/DRF-CNRS/IN2P3, Bvd Henri Becquerel, 14076 Caen, France}
\author{S.~Franchoo}
\affiliation{Université Paris-Saclay, CNRS, IJCLab, 91405, Orsay, France.}
\author{P.~Gangnant}
\affiliation{Grand Acc\'el\'erateur National d’Ions Lourds (GANIL), CEA/DRF-CNRS/IN2P3, Bvd Henri Becquerel, 14076 Caen, France}
\author{J.~Gibelin}
\affiliation{Universit\'{e} de Caen Normandie, ENSICAEN, CNRS/IN2P3, LPC Caen UMR6534, F-14000 Caen, France}
\author{J.~Goupil}
\affiliation{Grand Acc\'el\'erateur National d’Ions Lourds (GANIL), CEA/DRF-CNRS/IN2P3, Bvd Henri Becquerel, 14076 Caen, France}
\author{F.~Hammache}
\affiliation{Université Paris-Saclay, CNRS, IJCLab, 91405, Orsay, France.}
\author{C.~Houarner}
\affiliation{Grand Acc\'el\'erateur National d’Ions Lourds (GANIL), CEA/DRF-CNRS/IN2P3, Bvd Henri Becquerel, 14076 Caen, France}
\author{B.~Jacquot}
\affiliation{Grand Acc\'el\'erateur National d’Ions Lourds (GANIL), CEA/DRF-CNRS/IN2P3, Bvd Henri Becquerel, 14076 Caen, France}
\author{G.~Lebertre}
\affiliation{Grand Acc\'el\'erateur National d’Ions Lourds (GANIL), CEA/DRF-CNRS/IN2P3, Bvd Henri Becquerel, 14076 Caen, France}
\author{L.~Legeard}
\affiliation{Grand Acc\'el\'erateur National d’Ions Lourds (GANIL), CEA/DRF-CNRS/IN2P3, Bvd Henri Becquerel, 14076 Caen, France}
\author{L.~M\'{e}nager}
\affiliation{Grand Acc\'el\'erateur National d’Ions Lourds (GANIL), CEA/DRF-CNRS/IN2P3, Bvd Henri Becquerel, 14076 Caen, France}
\author{V.~Morel}
\affiliation{Grand Acc\'el\'erateur National d’Ions Lourds (GANIL), CEA/DRF-CNRS/IN2P3, Bvd Henri Becquerel, 14076 Caen, France}
\author{P.~Morfouace}
\affiliation{Université Paris-Saclay, CNRS, IJCLab, 91405, Orsay, France.}
\affiliation{CEA, DAM, DIF, F-91297 Arpajon, France.}
\author{J.~Pancin}
\affiliation{Grand Acc\'el\'erateur National d’Ions Lourds (GANIL), CEA/DRF-CNRS/IN2P3, Bvd Henri Becquerel, 14076 Caen, France}
\author{E.~C.~Pollacco}
\affiliation{Universit\'{e}  Paris-Saclay, IRFU, CEA, F-91191 Gif-sur-Yvette, France}
\author{M.~Rejmund}
\affiliation{Grand Acc\'el\'erateur National d’Ions Lourds (GANIL), CEA/DRF-CNRS/IN2P3, Bvd Henri Becquerel, 14076 Caen, France}
\author{T.~Roger}
\affiliation{Grand Acc\'el\'erateur National d’Ions Lourds (GANIL), CEA/DRF-CNRS/IN2P3, Bvd Henri Becquerel, 14076 Caen, France}
\author{F.~Saillant}
\affiliation{Grand Acc\'el\'erateur National d’Ions Lourds (GANIL), CEA/DRF-CNRS/IN2P3, Bvd Henri Becquerel, 14076 Caen, France}
\author{M.~S\'{e}noville}
\affiliation{Universit\'{e}  Paris-Saclay, IRFU, CEA, F-91191 Gif-sur-Yvette, France}

\date{\today}

\begin{abstract}
The  $^{18}$Ne(d,t)$^{17}$Ne and $^{18}$Ne(d,$^3$He)$^{17}$F single-nucleon pickup reactions were measured at 16.5\,MeV/nucleon in inverse kinematics together with elastic and inelastic scattering channels. The full set of measured exclusive differential cross sections was compared with the mirror reaction channels on stable $^{18}$O after consistent reanalysis using coupled reaction channels calculations. Within this interpretation scheme, most of the spectroscopic factors extracted for the population of unbound states in $^{17}$F match within uncertainties with their mirror partners in $^{17}$O. However, for the deeply-bound neutron removal channel to $^{17}$Ne, a significant symmetry breaking with the mirror proton-removal channel leading to $^{17}$N is evidenced by an overall single-particle strength reduction.\end{abstract}

\maketitle

\section{\label{Intro} Introduction}

Nucleon addition and removal reactions have been studied for many years with various probes and reaction mechanisms to examine the structure of atomic nuclei. Deviations from the independent particle model with evidence for short and long-range correlations were revealed in the pioneering $(e,e^\prime p)$ inelastic scattering experiments with electron beams and stable fixed targets~\cite{lap93}. Thanks to the major progress achieved in radioactive ion beam production, these studies have been extended to unstable nuclei using direct reaction measurements in inverse kinematics such as transfer, nucleon removal (knockout) or quasi-free scattering reactions. The extraction of spectroscopic information from the measured cross sections of these processes requires a good description of the reaction mechanism and nuclear structure inputs (overlap functions, densities, etc.) which are a source of possible inconsistencies when comparisons are made between these different probes (see for example the recent review concerning single-particle strength~\cite{aum21}).

In this perspective, several experimental efforts have focused on the study of $p$-shell nuclei by nucleon knockout at intermediate energies to benchmark the theoretical description of the reaction mechanism~\cite{baz09} or to compare the measured inclusive cross sections with those calculated using ab-initio structure model inputs~\cite{grin11}. When combined with recent exclusive measurements on the same systems~\cite{kuche}, the collected set of knockout cross sections allowed a detailed comparison for mirror pairs: for $i)$ $^{9}$Li/$^{8}$Li (2$^{+}$)  to $^{9}$C/$^{8}$B (2$^{+}$)~\cite{baz09} for weakly bound nucleon removal;
$ii)$ $^{10}$Be/$^{9}$Li (3/2$^{-}$)  to $^{10}$C/$^{9}$C (3/2$^{-}$)~\cite{grin11} at 120 MeV/u for deeply bound nucleon removal. When using Variational Monte-Carlo (VMC) structure inputs to calculate the theoretical cross sections the main conclusions are that similar spectroscopic factors (SFs) were obtained for mirror reactions but the reduction factors $R_s = \sigma_{exp} /\sigma_{th} $ obtained for deeply bound nucleon removal to the ground state moved significantly away from the global knockout systematics~\cite{tost} built using inclusive cross sections and shell model SFs. A dependence on the excitation energy of the final state was also observed, assigned in~\cite{kuche} to deficiencies in the calculation of nuclear structure data for excited states. \\

In general, it is assumed that mirror symmetry implies that SFs of mirror states should be identical. Note that in this work we define SF = $C^2S$, where $C$ is the {\em isospin} Clebsch-Gordan coefficient. For pairs of mirror states, as under consideration in this work, $C^2$ is identical and thus it is immaterial whether $C^2S$ or
$S$ is compared. However, for isobaric analog states $C^2$ is not the same for all members of a given multiplet and it is $S$ that should be compared in this case. To repeat, if mirror symmetry is assumed then both $C^2S$ and $S$ should be identical for pairs of mirror states. 

This assumption is often made for predicting astrophysically relevant cross sections using available information about mirror analogs. Within error bars, the nucleon knockout data on $p$-shell nuclei discussed above seem to confirm this assumption. However, several theoretical works predict that binding effects~\cite{Tim08} and coupling to the continuum~\cite{wyl21} could lead to differences between mirror SFs or asymptotic normalization coefficients (ANCs). Within a three-body model in~\cite{Tim08}, the authors found that such symmetry-breaking effects could arise at low proton-core binding energies. Conversely, in Ref.~\cite{wyl21} continuum-coupling effects are  predicted to be larger when deeply bound nucleons are removed because the daughter nucleus tends to move towards the dripline which leads to an appreciable change in configuration, thus reducing the overlap with the parent.\\

Nucleon transfer reactions at around 10 MeV/u provide an alternative mean for investigating these conclusions and predictions for mirror systems. Here we report on the $^{18}$Ne(d,t) and $^{18}$Ne(d,$^{3}$He) single nucleon pickup reactions measured in inverse kinematics with $^{18}$Ne projectiles incident on a deuterium target, with complete identification of final states. The measured differential cross sections are analysed by comparison with coupled reaction channel (CRC) calculations to quantify the $\left< ^{18}\mathrm{Ne} \mid \protect{^{17}\mathrm{Ne}} + n\right>$ and $\left< ^{18}\mathrm{Ne} \mid \protect{^{17}\mathrm{F}} + p\right>$ overlaps using the methodology described in~\cite{fla1,fla2}.
Existing pickup data obtained in direct kinematics at 26~MeV/nucleon with a deuteron beam incident on an $^{18}$O target~\cite{Har71,mair} are also analysed in a consistent fashion, enabling the comparison of pure mirror reactions for which the same spectroscopic factors may be expected.

\section{\label{Exp} Experiment}
 A $^{18}$Ne$^{10+}$ beam was produced and accelerated to 16.5\,MeV/nucleon by the SPIRAL facility at GANIL with a mean intensity of $1.2\,10^5$ pps. Two beam-tracking detectors (BTDs)~\cite{CATS} situated 1.2 m upstream of the secondary target were used to measure incoming beam trajectories event-by-event and determine the beam intensity for cross section normalization.

Depending on the reaction channel of interest, deuterated polypropylene (CD$_{2}$) targets of different thicknesses were used: 1.5\,mg.cm$^{-2}$ for elastic scattering, 3\,mg.cm$^{-2}$ for the (d,t) channels and 0.5\,mg.cm$^{-2}$ for the (d,$^{3}$He) channels. These targets were located at the target point of the VAMOS spectrometer~\cite{VAMOS}. Nominal target thicknesses were confirmed by dedicated measurements of the deviation of the unreacted beam in the VAMOS focal plane due to energy loss within the target. An uncertainty of 4\% in the target thickness resulted from this procedure.

The MUST2 array~\cite{MUST2} surrounded the target to detect and identify light charged particles in a configuration very similar to previous measurements~\cite{fla1,fla2}: four telescopes at forward angles to detect $^3$H and $^3$He and two at 90$^\circ$ relative to the beam axis for elastically scattered deuterons.  Identification of deuterons, tritons and $^3$He was achieved by $\Delta$E-E and $\Delta$E-TOF (Time-of-flight) techniques depending on the particle energy. Using a triple-alpha source ($^{239}$Pu, $^{241}$Am, $^{244}$Cm) for calibration,  an average energy resolution of 40(2)~keV (FWHM) was reached for the double-sided silicon strip detectors (DSSDs). By combining the beam trajectory determination from the BTDs and the detected hit position on the DSSDs, located at 15~cm from the target for the forward wall, the scattering angle resolution is approximately 0.4$^{\circ}$~(FWHM) in the laboratory frame.  

To reach an exclusive discrimination of each reaction channel, heavy ejectiles were also identified event-by-event in the focal plane of the VAMOS magnetic spectrometer~\cite{REJ11}. Two position sensitive drift chambers were used to measure the trajectory of the recoiling ions. Their energy loss ($\Delta$E) was measured in a segmented ionization chamber and their time of flight was measured between a plastic scintillator and the cyclotron radio frequency signal. Software reconstruction methods~\cite{REJ11,LEM23} were used to obtain magnetic rigidity and the path length of the ions. From these quantities, mass-over-charge and atomic number of the fragments were determined and used to select heavy transfer products directly ($^{17}$Ne/$^{17}$F) when their bound states were populated or their decay fragments when unbound states were reached. In summary, the overall identification methodology (charged particles in MUST2 and heavy reaction products in VAMOS) is identical to the one already detailed in~\cite{fla1,fla2}.

Simulations of the MUST2 array were performed using the \emph{nptool}~\cite{nptool} package to determine the detection efficiency profile including all experimental effects (beam profile, target thickness, survey positions, energy thresholds, missing strips, etc.). This profile was then used to correct experimental acceptance effects in the measured angular distributions and build the final differential cross sections shown in the following section (Fig.~\ref{fig:elastic_ang}, \ref{fig:dt_ang}, and \ref{fig:d3He_ang}).

\section{\label{Res} Results}

\subsection{Elastic and inelastic scattering}

\begin{figure}[t!]
\includegraphics[width=\columnwidth]{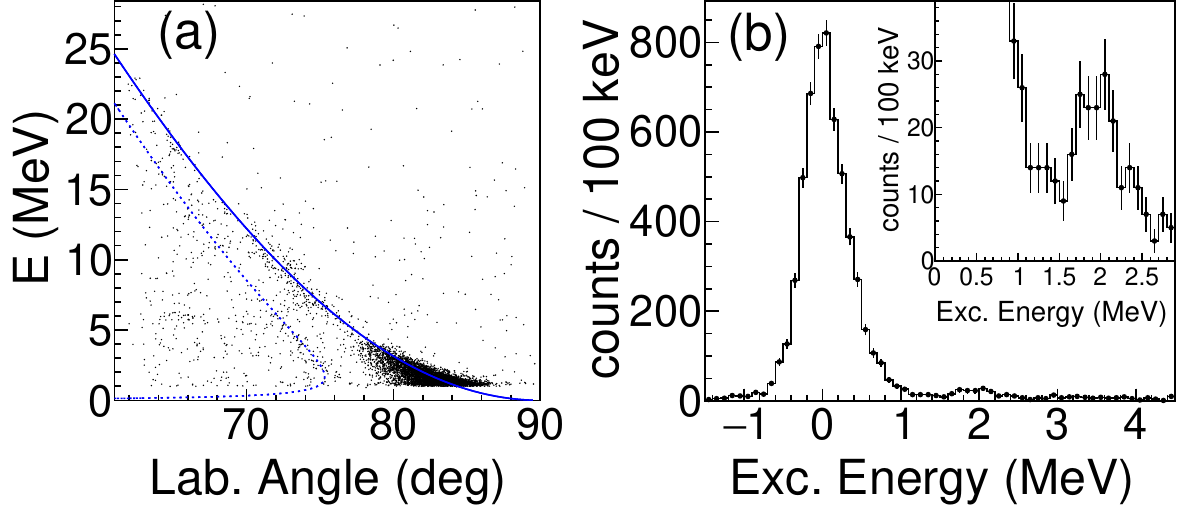}
\caption{\label{fig:elastic_kin} (a) Energy of the elastically scattered deuterons as a function of scattering angle in the laboratory frame. (b) Excitation energy spectrum of $^{18}$Ne.}
\end{figure}

\begin{figure}[t!]
\includegraphics[width=1\columnwidth]{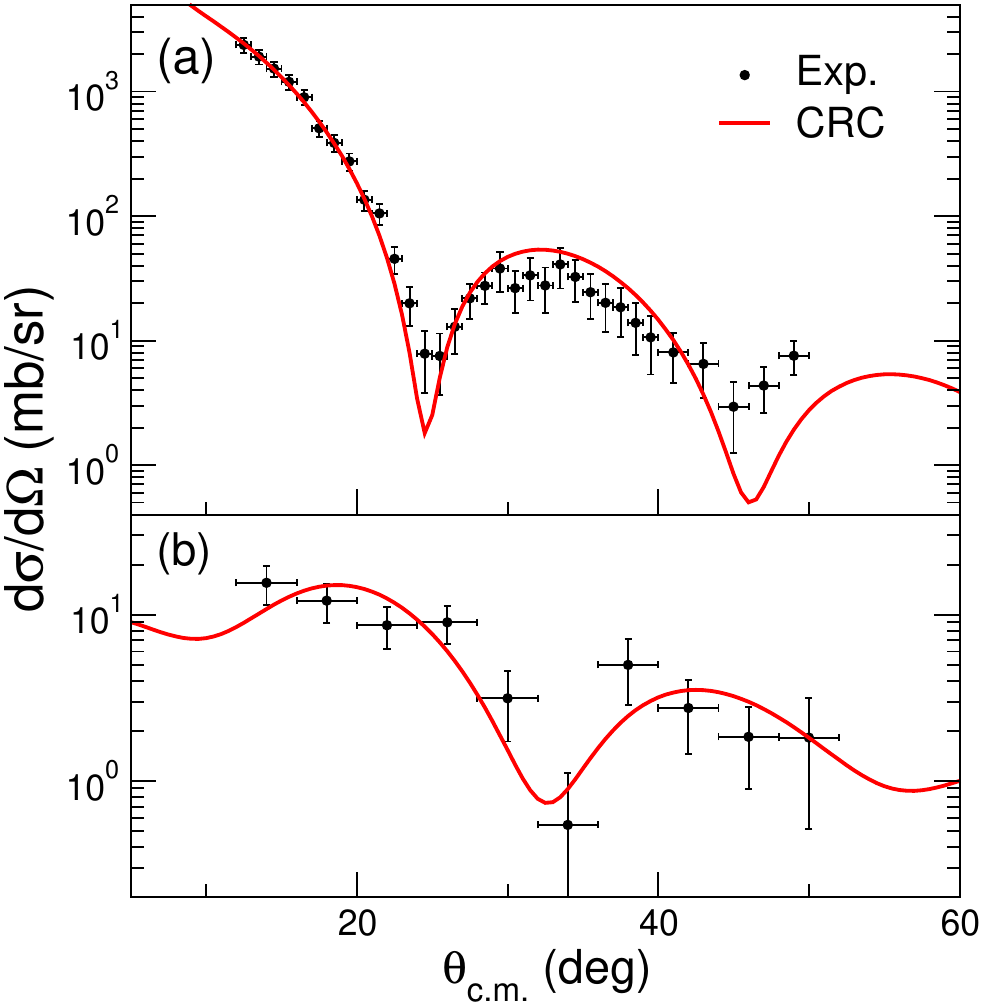}
\caption{\label{fig:elastic_ang} Angular distributions of (a)  $^{18}$Ne(d,d)$^{18}$Ne elastic scattering and (b) inelastic scattering to the $^{18}$Ne 2$^+_1$ state in the center of mass frame. Filled circles denote the experimental data with statistical and systematic uncertainties while the solid curves correspond to the results of a coupled reaction channels calculation.}
\end{figure}

The kinematics and excitation energy spectrum for the $^{18}$Ne(d,d) scattering are shown in Fig.~\ref{fig:elastic_kin} (a) and (b), respectively. A measured excitation energy resolution (FWHM) of 630~keV was obtained with the 1.5\,mg.cm$^{-2}$ thick target. Inelastic excitation of the $2^+$ state at 1887.3(2)\,keV is clearly observed, see the inset to Fig.~\ref{fig:elastic_kin} (b). The corresponding angular distributions are displayed in Fig.~\ref{fig:elastic_ang} (a) and (b), respectively.

\subsection{Neutron pickup $^{18}$Ne(d,t)$^{17}$Ne}

\begin{figure}[htpb!]
\includegraphics[width=1\columnwidth]{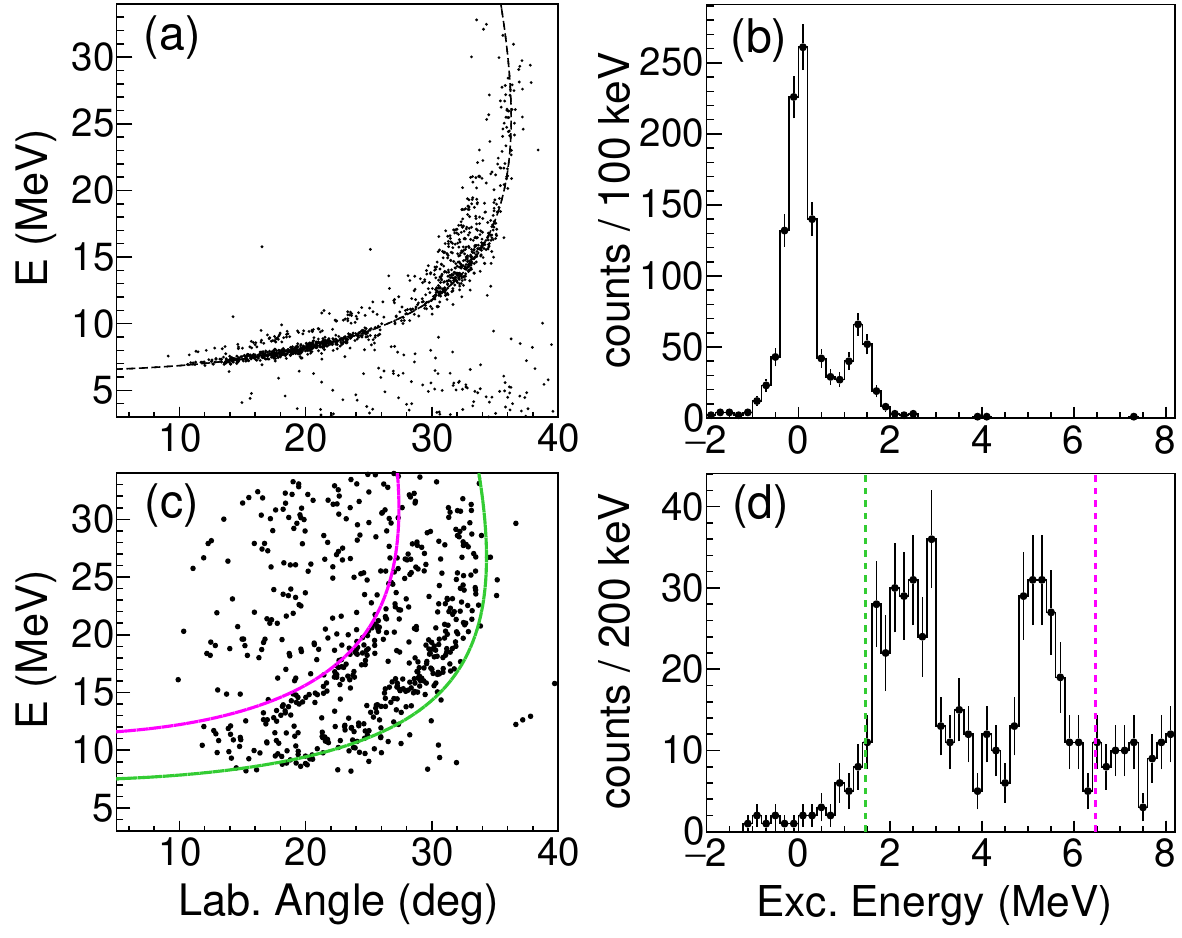}
\caption{\label{fig:dt_kin} Energy of tritons from the $^{18}$Ne(d,t)$^{17}$Ne reaction as a function of scattering angle in the laboratory frame and reconstructed excitation energy spectra for bound states (panels (a) and (b)) and proton unbound states (panels (c) and (d)).}
\end{figure}

The kinematics and excitation energy spectrum for the $^{18}$Ne(d,t)$^{17}$Ne reaction are shown in Fig.~\ref{fig:dt_kin}. The ground and first excited states, displayed in panels (a) and (b), were isolated by gating on a triton in MUST2 and a $^{17}$Ne residue in VAMOS. The proton-unbound states in panels (c) and (d) were selected by gating on a triton in MUST2 and a $^{15}$O residue in VAMOS since $^{16}$F is also proton unbound.

\begin{figure}[t!]
\includegraphics[width=1\columnwidth]{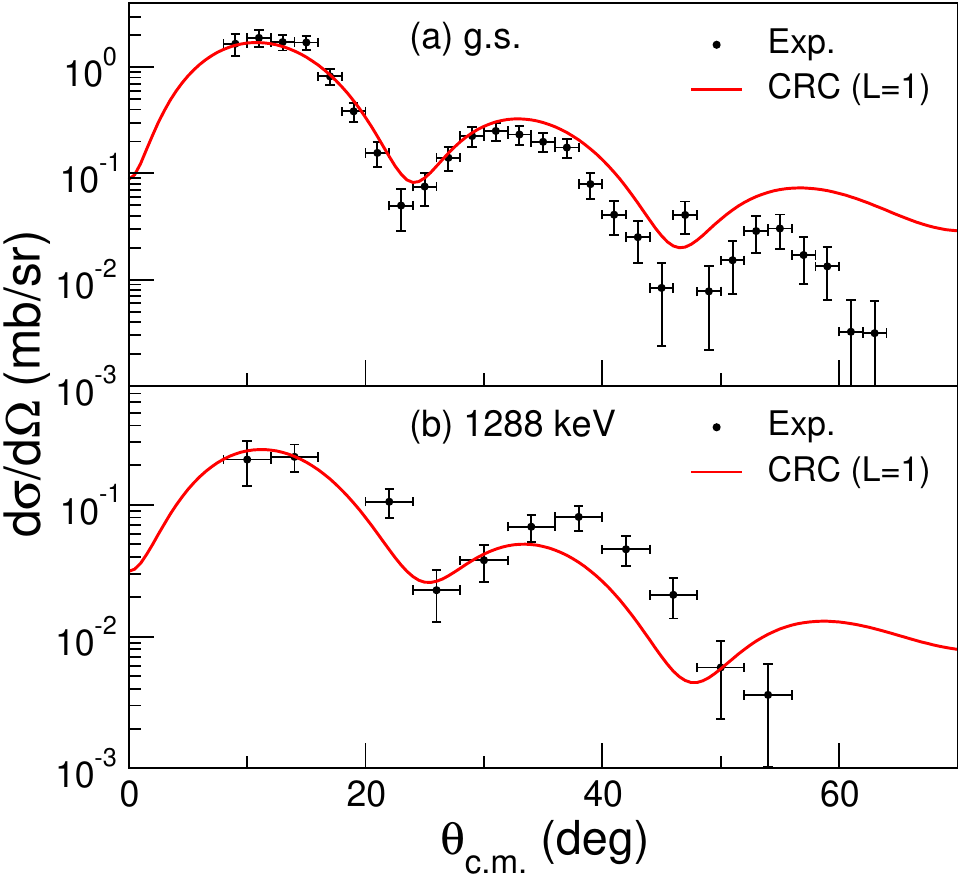}
\caption{\label{fig:dt_ang} Angular distributions for the $^{18}$Ne(d,t)$^{17}$Ne reaction to (a) the $1/2^-$ ground state and (b) the 1288 keV $3/2^-$ state. Filled circles denote the experimental data with statistical and systematic uncertainties while the solid curves correspond to the results of a coupled reaction channels calculation.}
\end{figure}

The ground and 1.288\,MeV first excited states of $^{17}$Ne, with spin-parities of $1/2^-$ and $3/2^-$, are well resolved and the shape of their angular distributions (Fig.\,\ref{fig:dt_ang}) matches with that expected for pickup of an L=1 neutron from the $1\,p_{1/2}$ and $1\,p_{3/2}$ valence orbitals of $^{18}$Ne, respectively.

For the unbound states the excitation energy spectra display a broad structure at 2.5 MeV and a peak around 5.5\,MeV. The width of the peak at 5.5 MeV is compatible with the simulated resolution for population of a single state, the mirror of the 5.52 MeV $3/2^-$ level in $^{17}$N, strongly populated in  the $^{18}$O(d,$^{3}$He) reaction \cite{Har71}. However, the width of the broad structure at 2.5 MeV requires it to contain at least two states.  Given this uncertainty and the rather poor statistics in this structure, angular distributions for these levels were not obtained.

\subsection{Proton pickup $^{18}$Ne(d,$^{3}$He)$^{17}$F}
\begin{figure}[t!]
\includegraphics[width=1\columnwidth]{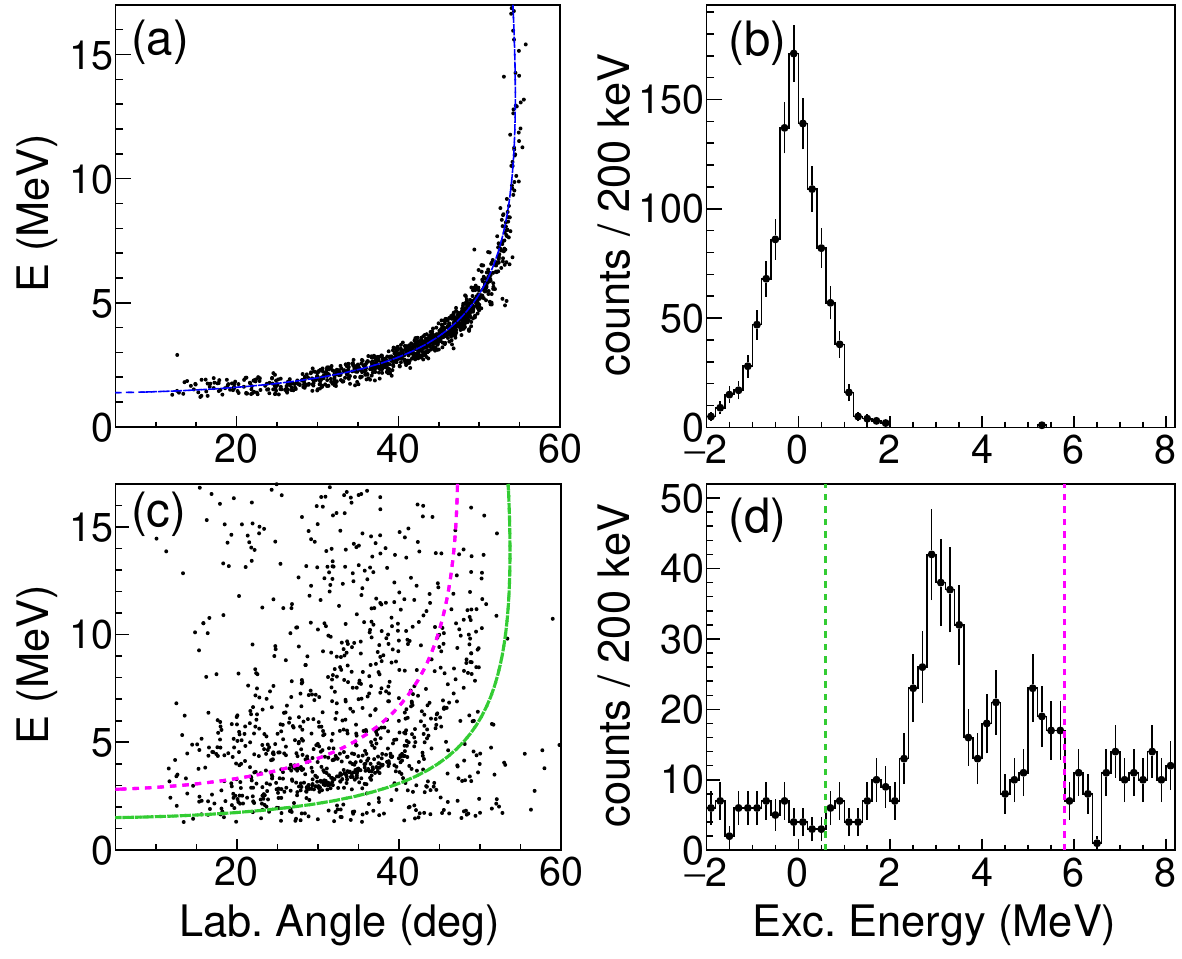}
\caption{\label{fig:d3He_kin} Energy of $^{3}$He from the $^{18}$Ne(d,$^{3}$He)$^{17}$F reaction as a function of scattering angle in the laboratory frame and reconstructed excitation energy spectrum for bound states (panels (a) and (b)) and proton unbound states (panels (c) and (d)).}
\end{figure}

\begin{figure}[t!]
\includegraphics[width=1\columnwidth]{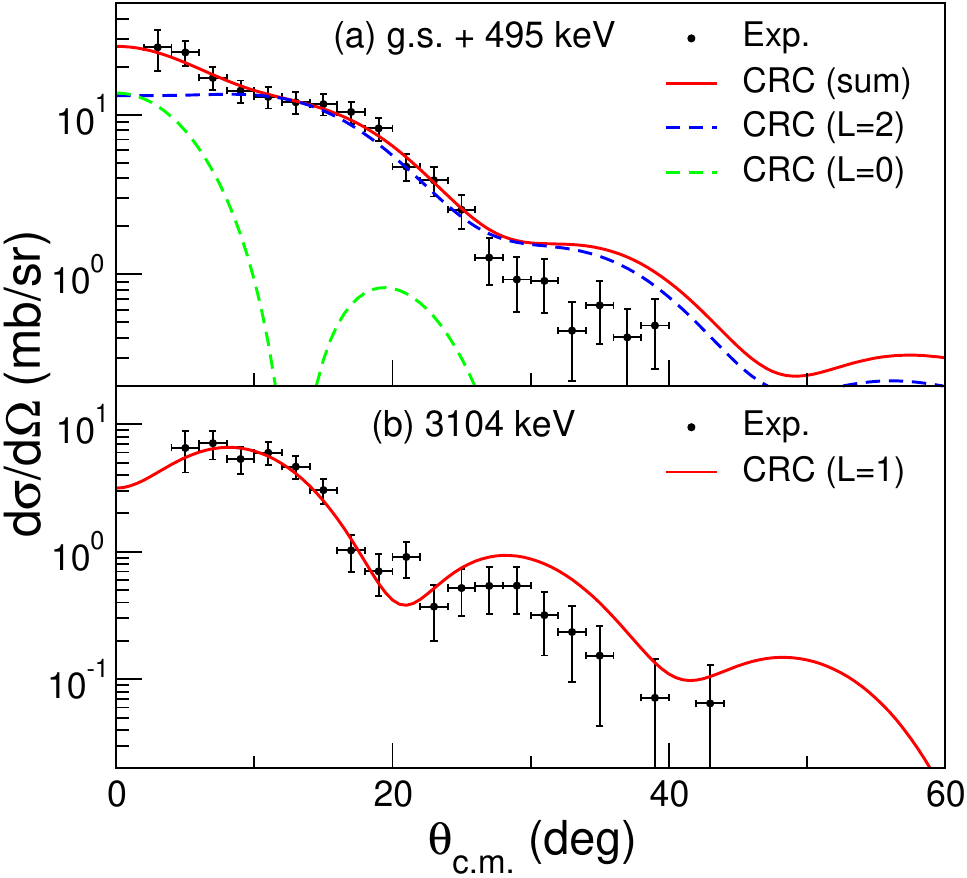}
\caption{\label{fig:d3He_ang} Angular distributions for the $^{18}$Ne(d,$^{3}$He)$^{17}$F reaction to (a) the unresolved $5/2^+$ ground state and 495 keV $1/2^+$ state (b) the 3104 keV $1/2^-$ state. Filled circles denote the experimental data with statistical and systematic uncertainties while the curves correspond to the results of a coupled reaction channels calculation.}
\end{figure}

The kinematics and excitation energy spectrum for the $^{18}$Ne(d,$^{3}$He)$^{17}$F reaction are shown in Fig.~\ref{fig:d3He_kin}. The $5/2^+$ ground and 495 keV $1/2^+$ first excited states of $^{17}$F, displayed in  Fig.~\ref{fig:d3He_kin} (a) and (b) and selected by gating on a $^{3}$He in MUST2 and a $^{17}$F residue in VAMOS, could not be resolved. However, their combined angular distribution shown on Fig.~\ref{fig:d3He_ang} exhibits the characteristic shape of an L=2 transfer, except for the two most forward angles, indicating that transfer to the $5/2^+$ ground state dominates. 

The kinematics and excitation energy spectrum for the proton unbound levels of $^{17}$F, obtained by gating on a  $^{3}$He in MUST2 and a $^{16}$O residue in VAMOS, are displayed in Fig.~\ref{fig:d3He_kin} (c) and (d). The statistics for populating these levels were such that an angular distribution was extracted for the 3104 keV $1/2^{-}$ level only.

\section{Analysis}

\subsection{$^{18}$Ne data}

The calculations were similar to those described in Refs.~\cite{fla1} and \cite{fla2}, employing a combination of the continuum discretized coupled channel (CDCC) technique to model the effects of deuteron breakup on the elastic scattering and the coupled reaction channels (CRC) formalism for the ($d$,$t$) and ($d$,$^3$He) transfer reactions. In addition, since data for the deuteron inelastic scattering to the 1.89-MeV $2_1^+$ excited state of $^{18}$Ne were also collected, coupling to this channel was included via the standard coupled channel (CC) method. All calculations were performed using the {\sc fresco} code \cite{Tho88}. 

The CDCC component was similar to that described in Ref.~\cite{Kee04}, with the continuum divided into momentum ($k$) bins of width $\Delta k = 0.125$ fm$^{-1}$ up to a maximum value of $k_\mathrm{max} = 0.75$ fm$^{-1}$. The entrance channel deuteron potentials, including all coupling potentials, were constructed by Watanabe folding of the central parts of global nucleon optical potentials over the deuteron internal wave function, the latter being calculated with the Reid soft core nucleon-nucleon potential \cite{Rei68}. Calculations were performed with deuteron potentials based on four different global nucleon parameter sets, {\it viz.}: those of Becchetti and Greenlees \cite{Bec69}, Watson {\it et al.\/} \cite{Wat69}, Varner {\it et al.\/} (CH89) \cite{Var91}, and Koning and Delaroche~\cite{Kon03}. The real and imaginary depths of the Watanabe potentials (including the coupling potentials) were multiplied by factors $\lambda_V$ and $\lambda_W$ to give the best description of the measured elastic scattering angular distribution. The values of $\lambda_V$ and $\lambda_W$ varied from 1.0 by at most $\pm 15$\% and $+20$\%, respectively. The strength of
the Coulomb coupling to the 1.89-MeV $2_1^+$ excited state of $^{18}$Ne was fixed using the recommended value of the $B(E2)$ from Pritychenko {\it et al.\/} \cite{Pri16}. The nuclear coupling was included by deforming the diagonal Watanabe folding potential for the entrance channel and adjusting the nuclear deformation length to give the best agreement with the measured inelastic scattering angular distribution. This resulted in a deformation length $\delta = 1.36$ fm.

The transfer steps included the full complex remnant terms and non-orthogonality corrections. Couplings were included for pickup leading to the 0.0-MeV $1/2^-$ and 1.35-MeV $3/2^-$ levels of $^{17}$O and the 0.0-MeV $5/2^+$, 0.4953-MeV $1/2^+$, and 3.104-MeV $1/2^-$ levels of $^{17}$F. In the absence of suitable
elastic scattering data global parameter sets were employed for the exit channel $t$ + $^{17}$Ne and $^3$He + $^{17}$F optical potentials. Calculations were performed using the mass-three parameter sets of Refs.\ \cite{Pan09} and \cite{Pan15}; the latter is slightly outside its region of applicability for these systems, being specifically adapted for scattering from $1p$-shell targets. The $\left<t \mid d + n \right>$ and $\left< ^3\mathrm{He} \mid d + p \right>$ overlaps were calculated following the procedure of Brida {\it et al.\/} \cite{Bri11}. The $\left<^{18}\mathrm{Ne} \mid \protect{^{17}\mathrm{Ne}} + n \right>$ and $\left<^{18}\mathrm{Ne} \mid \protect{^{17}\mathrm{F}} + p \right>$ overlaps were calculated using the methodology described in Refs.\ \cite{fla1} and \cite{fla2}. For each level of the residual nuclei under consideration the transferred neutron or proton was bound in a Woods-Saxon well with a diffuseness of 0.65 fm and a radius adjusted so that the r.m.s.\ radius of the wave function reproduced that of the corresponding shell model orbital obtained from a Hartree-Fock-Bogoliubov (HFB) calculation with the Sly4 interaction~\cite{HFBrad,Sly4}. The values are given in Table~\ref{tab1}. The binding potential also included a spin-orbit term of Thomas form with a fixed depth of 9.0 MeV. The depth of the central Woods-Saxon well was adjusted to reproduce the appropriate binding energy. The experimental spectroscopic factors $\mathrm{C^2S}_{exp}$ were obtained by normalizing the calculated angular distributions to the corresponding transfer data. The mean values of $\mathrm{C^2S}_{exp}$ for the eight different combinations of entrance and exit channel optical potentials are given in Table \ref{tab1}. Although not statistically rigorous due to the small sample size, 95\% confidence limits calculated from the sample standard deviations are also listed in the table in the form of ``error bars'' as an indication of the spread of values extracted from the individual calculations.

The results of a typical calculation (that with entrance channel potentials based on the 
Koning and Delaroche (KD) global nucleon optical potential \cite{Kon03} and the global mass-three optical potential (A=3 OP) of Pang {\it et al.}~\cite{Pan09}) are compared with the $^{18}$Ne + d data in Figs.~\ref{fig:elastic_ang}, \ref{fig:dt_ang} and \ref{fig:d3He_ang}.

\subsection{$^{18}$O data}

The calculations for the existing $^{18}$O($d$,$t$)$^{17}$O and $^{18}$O($d$,$^3$He)$^{17}$N data were identical in methodology to the $d$ + $^{18}$Ne calculations described above. However, due to the greater incident energy the continuum space of the CDCC component was increased to $k_\mathrm{max} = 1.0$ fm$^{-1}$. Since there are no $d$ + $^{18}$O elastic scattering data available at the appropriate energy the $\lambda_V$ and $\lambda_W$ values were obtained by adjusting to fit the 52-MeV $d$ + $^{16}$O elastic scattering data of Hinterberger {\it et al.}~\cite{Hin68}. The values of $\lambda_V$ and $\lambda_W$ varied from 1.0 by at most $\pm 10$\% and $-25$\%, respectively. Coupling to the 1.98-MeV $2_1^+$ level of $^{18}$O was included in an analogous fashion to that to the 1.89-MeV $2_1^+$ level of $^{18}$Ne, with the Coulomb coupling strength fixed using the recommended $B(E2)$ value of Raman {\it et al.}~\cite{Ram01}. Since no inelastic scattering data are available the nuclear deformation length was fixed at the value obtained from the $B(E2)$ assuming the collective model and a charge radius of $1.2 \times 18^{1/3}$ fm, giving $\delta = 1.12$ fm.

\begin{figure}
\includegraphics[width=\columnwidth]{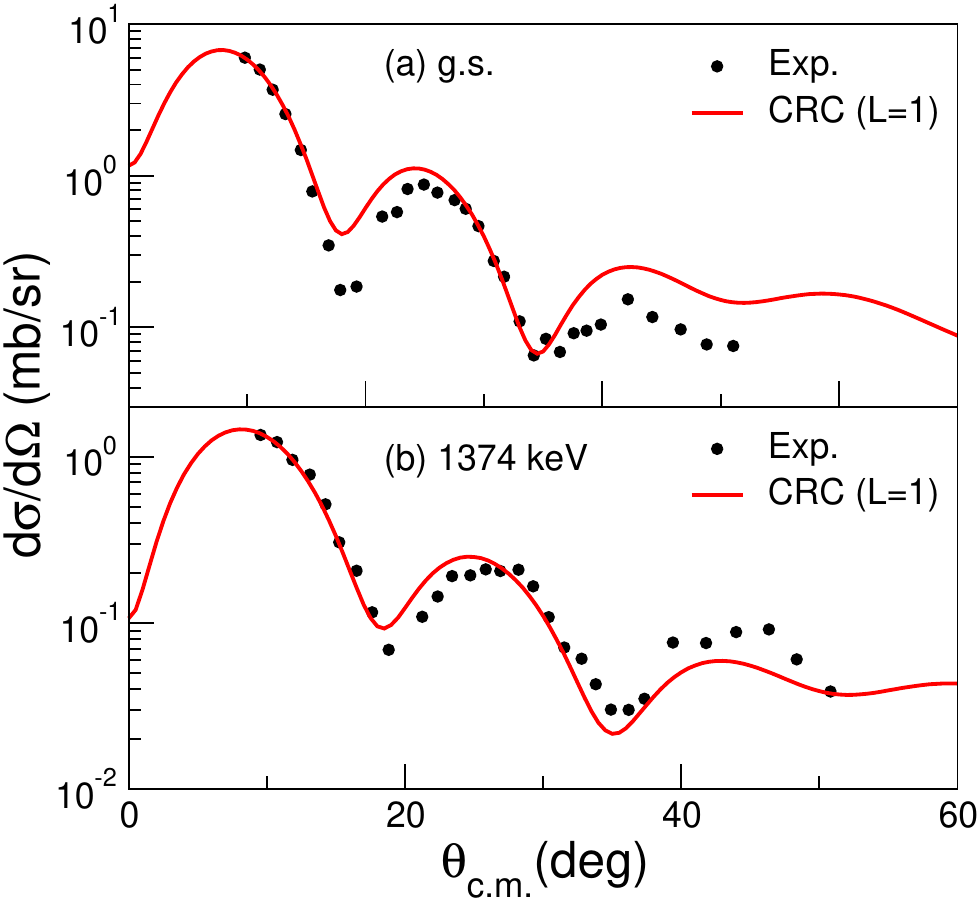}
\caption{\label{fig:18Od3He} Angular distributions for the $^{18}$O(d,$^{3}$He)$^{17}$N reaction to (a) the $1/2^-$ ground state (b) the 1374 keV $3/2^-$ state. Filled circles denote the experimental data of Ref.~\cite{Har71} while the curves correspond to the results of a coupled reaction channels calculation.}
\end{figure}

\begin{figure}
\includegraphics[width=\columnwidth]{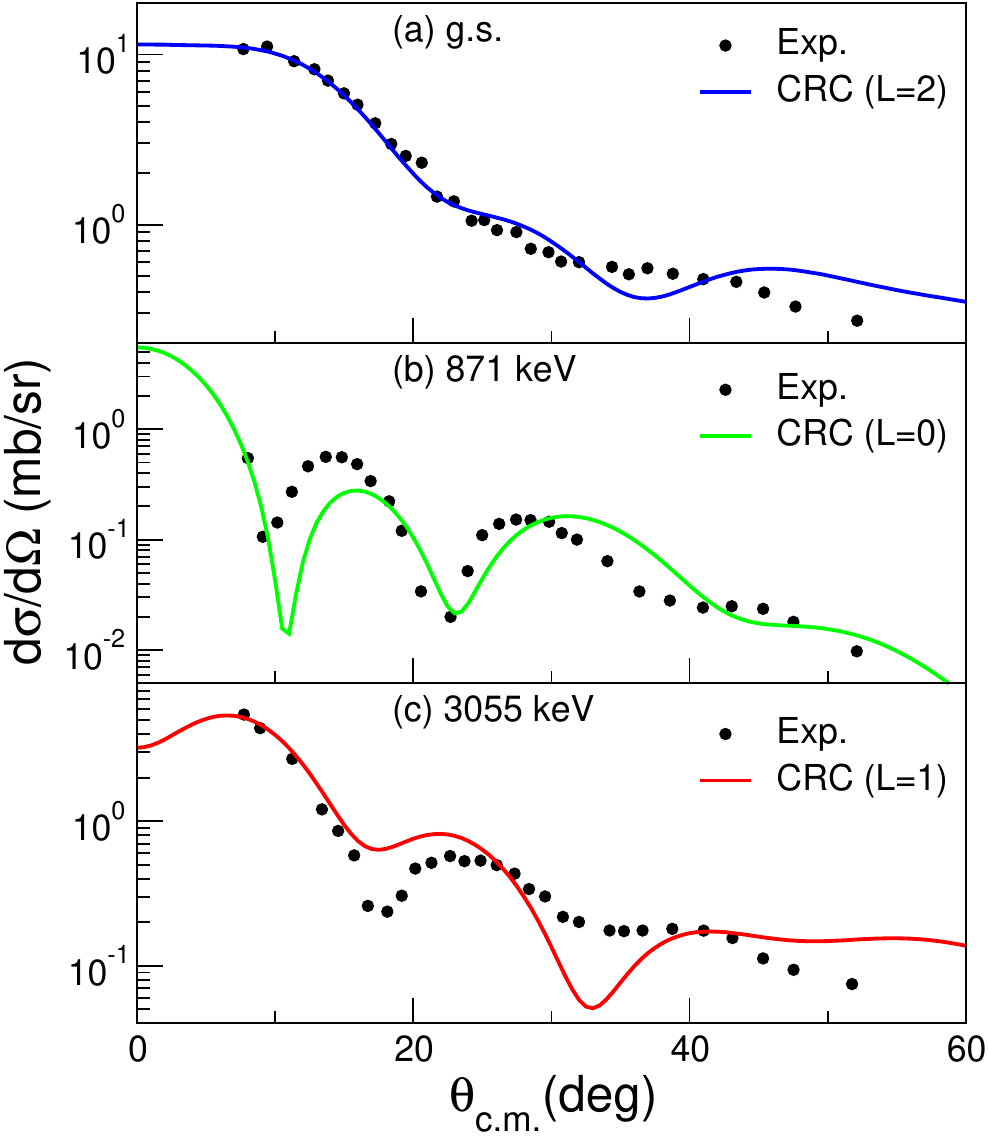}
\caption{\label{fig:18Odt} Angular distributions for the $^{18}$O(d,t)$^{17}$O reaction to (a) the $5/2^+$ ground state, (b) the 871 keV $1/2^+$ state and (c) the 3055 keV $1/2^-$ state. Filled circles denote the experimental data of Ref.~\cite{mair} while the curves correspond to the results of a coupled reaction channels calculation.}
\end{figure}

The transfer steps employed the same methodology as those described previously for the $d$ + $^{18}$Ne data. Couplings were limited to pickup leading to the $5/2^+$ ground state, 871 keV $1/2^+$, and 3055 keV $1/2^-$ levels of $^{17}$O and the $1/2^-$ ground state and 1374 keV $3/2^-$ level of $^{17}$N in order to match the $d$ + $^{18}$Ne calculations as closely as possible. Mean $\mathrm{C^2S}$ values, together with their 95\% confidence limits, are given in Table \ref{tab1}. The results of a typical calculation (that with entrance (KD)~\cite{Kon03} and exit (A=3 OP)~\cite{Pan09} channels potentials) are compared with the $^{18}$O(d,$^3$He) and $^{18}$O(d,t) data of Refs.~\cite{Har71} and \cite{mair}, respectively in Figs.~\ref{fig:18Od3He} and \ref{fig:18Odt}.


\section{\label{Int} Discussion}

The single nucleon pickup data for both $^{18}$Ne and $^{18}$O are in general well described by the CRC calculations, see Figs.~\ref{fig:dt_ang}, \ref{fig:d3He_ang}, \ref{fig:18Od3He}, and \ref{fig:18Odt},  as are the $^{18}$Ne + d elastic and inelastic scattering data, cf.\ Fig.~\ref{fig:elastic_ang}. This supports the assumption that the dominant mechanism for these reactions is one-step nucleon pickup. The only exceptions are the 495 keV $1/2^+$ level of $^{17}$F and its mirror partner the 871 keV $1/2^+$ level of $^{17}$N. The former could not be separated from the $5/2^+$ ground state of $^{17}$F, thus the value obtained for the SF for this level rests mainly on the fit to the first two points of the combined angular distribution, see Fig.~\ref{fig:d3He_ang} (a). The shapes of the angular distributions for the L = 2 transfer to the $5/2^+$ ground state and the L = 0 transfer to the 495 keV $1/2^+$ are significantly different, and since the shape of the measured combined angular distribution is clearly dominated by L = 2 transfer to the $5/2^+$ ground state, the extraction of a SF for this level was not significantly affected by the presence of the unseparated $1/2^+$. Conversely, the value of the SF for the 495 keV $1/2^+$ level given in Table \ref{tab1} should be considered an upper limit. For the 871 keV $1/2^+$ level of $^{17}$O the calculations do not well reproduce the positions of the minima of the measured angular distribution, being offset by several degrees, which will add to the uncertainty of the extracted SF. This is a relatively common phenomenon for L = 0 transfers and the original analysis exhibited the same problem, see Fig. 2 of  Mairle {\em et al.}~\cite{mair}. It may be indicative of a significant contribution to the population of this level from two-step reaction mechanisms which could significantly impact the obtained SF. It is not possible to determine from the available data whether the calculated angular distribution for the 495 keV $1/2^+$ level of $^{17}$F suffers from a similar problem. However, test calculations including the $\left<^{18}\mathrm{Ne}(2^+_1) \mid \protect{^{17}\mathrm{F}(1/2^+_1} + p\right>$ or $\left<^{18}\mathrm{O}(2^+_1) \mid \protect{^{17}\mathrm{O}(1/2^+_1} + n\right>$ overlaps (assuming a $2^+ \otimes d_{5/2}$ configuration) were able to describe satisfactorily the corresponding $^{18}$Ne(d,$^3$He) and $^{18}$O(d,t) angular distributions with the same SFs for both systems, i.e., complete fulfillment of mirror symmetry.

\begin{center}
\begin{table*}[t!]
\begin{tabular*}{\textwidth}{@{\extracolsep{\fill}} ccccc|cccc|ccc } 
\hline
\hline
     &   &   &  &  & \multicolumn{4}{c|}{HFB constrained $r_{0}$ }  & \multicolumn{3}{c}{$r_{0}=1.25$\,fm}\\
\hline Beam     & Reaction     &  Residue   & E      & $J^{\pi}$ & $r^{HF}_{rms}$& $r_0$   & $C^2S_{exp}$ & $ANC^{2}_{exp}$ & $r_{rms}$ & $C^2S_{exp}$ & $ANC^2_{exp}$\\
& & & (MeV) & &(fm) &(fm) & & & (fm) & \\
\hline $^{18}$Ne &   (d,t)     & $^{17}$Ne  &   0.0  & 1/2$^{-}$ &   2.893     &   1.480   &  $0.72(9)$ & $179(21)$ & 2.627  &  $1.28(15)$ & $160(19)$ \\ 
                 &             &            &  1.288 & 3/2$^{-}$ &   2.794     &   1.326   &  $0.22(3)$ & $55(8)$ & 2.699  &  $0.27(3)$ & $52.3(73)$ \\ 

\hline $^{18}$O & (d,$^{3}$He) & $^{17}$N   &   0.0  & 1/2$^{-}$ &  2.915     &   1.465   &  $1.15(16)$ & $440(61)$ & 2.662 &$1.97(28)$ & $402(58)$ \\ 
                &              &            &  1.374 & 3/2$^{-}$ &   2.811     &   1.311  &  $0.37(6)$ & $146(22)$ & 2.734 &$0.44(7)$ & $142(22)$ \\ 
\hline
\hline
$^{18}$Ne &(d,$^{3}$He) & $^{17}$F   &   0.0  & 5/2$^{+}$ &   3.409     &   1.244   &  $1.04(10)$  & $8.1(8)$ & 3.419  & $1.04(10)$ & $8.2(8)$\\ 
                &              &            &  0.495 & 1/2$^{+}$ &   3.568     &   1.108   &  $0.14(1)$  &  $15.3(12)$ & 3.716  & $0.12(1)$ & $15.5(12)$\\ 
                &              &            &  3.104 & 1/2$^{-}$ &   2.917     &   1.214   &  $0.55(5)$ & $27.3(27)$ & 2.959  & $0.52(5)$ & $27.9(26)$\\

\hline $^{18}$O &   (d,t)      & $^{17}$O   &   0.0  & 5/2$^{+}$ &  3.340     &   1.272   &  $1.04(12)$ & $5.9(7)$ & 3.307 & $1.08(13)$ & $5.8(7)$ \\ 
                &              &            &  0.871 & 1/2$^{+}$ &   3.443     &   1.210   &  $0.08(1)$ & $5.4(7)$ & 3.482 & $0.08(1)$  & $5.8(7)$ \\ 
                &              &            &  3.055 & 1/2$^{-}$ &   2.891     &   1.267   &  $0.61(6)$ & $16.6(17)$ & 2.872 & $0.62(7)$  & $16.3(18)$ \\
\hline
\hline
\end{tabular*}
 \caption{Mean spectroscopic factors ($\mathrm{C^2S_{exp}}$) and squared asymptotic normalization coefficients ($\mathrm{ANC^2_{exp}}$) for single nucleon removal from $^{18}$Ne and $^{18}$O obtained in this work from CRC calculations for two different prescriptions concerning the radius parameter of the Woods-Saxon potential binding the transferred nucleon to the target-like core ($r_0$): (i) adjusted to reproduce the rms radii of the corresponding orbitals obtained from HFB calculations with the SLy4 interaction~\cite{HFBrad,Sly4} or (ii) a standard fixed value of $1.25$\,fm.}
 \label{tab1}
 \end{table*}
 \end{center}

The most significant result of the present analysis concerns the mirror symmetry, whereby the SFs for the $\left<\mathrm{^A_ZX_N} \mid \protect{\mathrm{^{A-1}_{Z-1}Y_N}} + p \right>$ and $\left<\mathrm{^A_NV_Z} \mid \protect{\mathrm{^{A-1}_{\;\;\;\;N}W_{Z-1}}} + n \right>$ mirror overlaps should be identical. This appears to hold good for the $\left<^{18}\mathrm{Ne} \mid \protect{^{17}\mathrm{F}} + p \right>$ and $\left<^{18}\mathrm{O} \mid \protect{^{17}\mathrm{O}} + n \right>$  overlaps, at least for the $5/2^+$ and $1/2^-$ states, cf.\ the ratios of the spectroscopic factors $R^\mathrm{M}_\mathrm{HF}$ listed in Table \ref{tab4}. It is not possible to draw definite conclusions concerning the factor of $\approx 2$ difference between the values for the $1/2^+$ states due to the difficulties in obtaining well defined SFs for these levels discussed in the previous paragraph (the uncertainties in the SFs for these levels listed in Table \ref{tab1} do not take these into account). In contrast, the SFs for the $\left<^{18}\mathrm{Ne} \mid \protect{^{17}\mathrm{Ne}} + n \right>$ and $\left<^{18}\mathrm{O} \mid \protect{^{17}\mathrm{N}} + p \right>$ overlaps appear to violate mirror symmetry, with $R^\mathrm{M}_\mathrm{HF}$ values of $\approx 0.6$ for both the $1/2^-$ and $3/2^-$ levels, significantly different from unity.

However, it may be argued that this apparent breaking of mirror symmetry is simply due to the particular choice of binding potential radius parameter values. If we adopt the frequently employed methodology where the radius parameters of the Woods-Saxon potentials binding the transferred nucleons to the target-like residual nuclei are fixed at $r_0 = 1.25 \times 17^{1/3}$ fm for all states but retain all other inputs unchanged, a similar picture nevertheless emerges. The SFs for the $\left<^{18}\mathrm{Ne} \mid \protect{^{17}\mathrm{Ne}} + n \right>$ and $\left<^{18}\mathrm{O} \mid \protect{^{17}\mathrm{N}} + p \right>$ overlaps continue to violate mirror symmetry by about 40\% while those for the $\left<^{18}\mathrm{Ne} \mid \protect{^{17}\mathrm{F}} + p \right>$ and $\left<^{18}\mathrm{O} \mid \protect{^{17}\mathrm{O}} + n \right>$ overlaps agree well, see the ratios $R^\mathrm{M}_{1.25}$ in Table \ref{tab4} (we again exclude the $1/2^+$ states from this comparison). The SFs for the $\left<^{18}\mathrm{Ne} \mid \protect{^{17}\mathrm{Ne}} + n\right>$ and $\left<^{18}\mathrm{O} \mid \protect{^{17}\mathrm{N}} + p\right>$ overlaps are significantly different from those obtained with the tuned $r_0$ values, but this merely reflects the well known dependence on the binding potential radius (with the exception of the 0.495-MeV $1/2^+$ level of $^{17}$F all the other ``tuned'' $r_0$ values listed in Table \ref{tab1} are close to 1.25).

\begin{center}
\begin{table*}[t!]
\begin{tabular*}{\textwidth}{@{\extracolsep{\fill}} cccccc } 
\hline
\hline Mirror pair & $J^{\pi}$ & $R^\mathrm{M}_\mathrm{HF}$ & $R^\mathrm{M}_{1.25}$ & $R^\mathrm{M}(\mathrm{ANC^2})_\mathrm{HF}$ & $R^\mathrm{M}(\mathrm{ANC^2})_{1.25}$ \\
\hline $^{17}$Ne : $^{17}$N  & 1/2$^{-}$ &  $0.63(12)$    &  $0.65(12)$ & $0.62(11)$ & $0.65(12)$\\ 
                             & 3/2$^{-}$ &  $0.60(13)$    &  $0.61(12)$ & $0.59(12)$ & $0.61(13)$\\ 
\hline
\hline $^{17}$F :  $^{17}$O  & 5/2$^{+}$ &  $1.00(15)$    &  $0.96(15)$ & $1.01(15)$ & $0.97(15)$\\ 
                             & 1/2$^{+}$ &  $1.75(25)$    &  $1.50(23)$ & $1.72(25)$ & $1.39(20)$\\ 
                             & 1/2$^{-}$ &  $0.90(12)$    &  $0.84(12)$ & $0.91(13)$ & $0.84(12)$\\ 
\hline
\hline
\end{tabular*}
 \caption{Ratios of spectroscopic factors for mirror pairs extracted from the CRC calculations with $r_0$ values tuned to reproduce the HF rms radii ($R^\mathrm{M}_\mathrm{HF}$) and with $r_0 = 1.25$ fm for all levels ($R^\mathrm{M}_{1.25}$) together with the corresponding ratios for the $\mathrm{ANC^2}$ values obtained using Eq.~\ref{eq1}.}
\label{tab4}
\end{table*}
\end{center}

Although both sets of calculations, those with the tuned $r_0$ values and those with $r_0 = 1.25$ fm for all target-like overlaps, lead to the same conclusions concerning the violation of mirror symmetry in these systems (the ratios $R^\mathrm{M}_\mathrm{HF}$ and $R^\mathrm{M}_{1.25}$ are in excellent quantitative agreement) one would ideally like to find some measure which does not depend on the choice of binding potential radius. The asymptotic normalization coefficient, ANC, is usually much less influenced by the choice of binding potential parameters than the SF, and is normally essentially independent of $r_0$ over quite large ranges of values. It can thus provide us with just such a measure as we require. If mirror symmetry holds then, to a good approximation, the ANCs for mirror pair levels should satisfy the following relation \cite{Tra03}:
\begin{equation}\label{eq1}
    C^2_{\ell j}(\mathrm{^A_Z X_N}) = C^2_{\ell j}(\mathrm{^A_NY_Z)} b^2_{\ell j}(\mathrm{^A_Z X_N})/b^2_{\ell j}(\mathrm{^A_NY_Z)}
\end{equation}
where the $C^2_{\ell j}$ are the square ANCs for the respective proton and neutron overlaps where the transferred nucleon has angular momentum and total spin $\ell$ and $j$, respectively and the $ b^2_{\ell j}$ are the corresponding squared single-particle ANCs. Thus, the ratio of the right-hand side of Eq.~(\ref{eq1}) to the left-hand side, which we shall denote $R^\mathrm{M}(\mathrm{ANC^2})$,  provides a measure of the degree to which mirror symmetry is satisfied. The values for both sets of calculations are presented in Table \ref{tab4} and are in good agreement not only with each other but also the corresponding SF ratios.

Our analysis thus leads to the strong conclusion that while mirror symmetry holds well for the $\left< ^{18}\mathrm{Ne} \mid \protect{^{17}\mathrm{F}} + p\right>$ and $\left< ^{18}\mathrm{O} \mid \protect{^{17}\mathrm{O}} + n\right>$ mirror overlaps (with the exception of the $1/2^+_1$ levels of $^{17}$F and $^{17}$O for which no definite conclusions may be drawn) it is significantly broken for the $\left< ^{18}\mathrm{Ne} \mid \protect{^{17}\mathrm{Ne}} + n\right>$ and $\left< ^{18}\mathrm{O} \mid \protect{^{17}\mathrm{N}} + p\right>$ mirror overlaps. Our results for the $\left< ^{18}\mathrm{Ne} \mid \protect{^{17}\mathrm{F}} + p\right>$ and $\left< ^{18}\mathrm{O} \mid \protect{^{17}\mathrm{O}} + n\right>$ overlaps where the cores are in the $5/2^+$ ground states are consistent with the conclusions of Ref.~\cite{Tim08}. In that work the influence of three-body structures on mirror SFs and ANCs was examined and it was found that ``the possibility of symmetry breaking in mirror spectroscopic factors \ldots becomes more important at low p-core binding energies. It arises because at low p-core binding energies the proton spectroscopic factors are influenced by threshold effects, whereas in the mirror system the n-core energy is always larger, so that the mirror neutron spectroscopic factor is not influenced by the near threshold effects". However, for realistic N-core binding energies the mirror symmetry breaking for $\left< \mathrm{core} + N \mid \mathrm{core} + N + N \right>$ overlaps did not exceed approximately 5\% for large components such as the $5/2^+ \otimes d_{5/2}$ in the $0^+_1$ ground states of $^{18}$Ne and $^{18}$O. For the small $1/2^+ \otimes s_{1/2}$ components, corresponding to the cores in their $1/2^+_1$ excited states, the symmetry breaking was predicted to reach up to 25\%, although unfortunately were are unable to confirm this.

Notwithstanding, our most interesting conclusion concerns the $\left< ^{18}\mathrm{Ne} \mid \protect{^{17}\mathrm{Ne}} + n\right>$ and $\left< ^{18}\mathrm{O} \mid \protect{^{17}\mathrm{N}} + p\right>$ mirror overlaps. Our results clearly show a significant breaking of the spectroscopic factor mirror symmetry for these systems, significantly larger than the 25\%, described as ``unusually large,'' predicted in Ref.~\cite{Tim08} for the $1/2^+ \otimes s_{1/2}$ components in the $0^+_1$ ground states of $^{18}$Ne and $^{18}$O, despite the fact that $S_p = 13.1$ MeV for $^{17}$N and $S_n = 15.6$ MeV for $^{17}$Ne. It is therefore evident that in this case three-body threshold effects cannot provide an explanation and other, stronger influences must be at work. For example, continuum-coupling effects investigated in~\cite{wyl21} were predicted to be larger when deeply-bound nucleons are involved and dedicated calculations in this direction for the $\left< ^{18}\mathrm{Ne} \mid \protect{^{17}\mathrm{Ne}} + n\right>$ and $\left< ^{18}\mathrm{O} \mid \protect{^{17}\mathrm{N}} + p\right>$ mirror overlaps are called for to determine if they could lead to the symmetry breaking inferred here.\\
 
\begin{acknowledgments}
The authors would like to thank the SPIRAL beam delivery team and the VAMOS  technical
staff for their excellent support.
This work was supported by LIA COPIGAL.

\end{acknowledgments}

\end{document}